\documentclass[prd,aps,showpacs,nofootinbib]{revtex4}%
\usepackage{graphicx}
\usepackage{amsmath}
\usepackage{amsfonts}
\usepackage{amssymb}%
\newcommand{\be}{\begin{equation}}
\newcommand{\ee}{\end{equation}}
\newcommand{\bq}{\begin{eqnarray}}
\newcommand{\eq}{\end{eqnarray}}
\begin{document}
\title{\textbf{Effective potential and spontaneous symmetry breaking in the
noncommutative }$\varphi^{6}$\textbf{\ model}}
\author{G. D. Barbosa}
\email{gbarbosa@cbpf.br}
\affiliation{Centro Brasileiro de Pesquisas F\'{\i}sicas, CBPF, Rua Dr. Xavier Sigaud 150,
22290-180, Rio de Janeiro, Brazil }

\begin{abstract}
We study the conditions for spontaneous symmetry breaking of the
(2+1)-dimensional noncommutative $\varphi^{6}$ model in the small-$\theta$
limit. In this regime, considering the model as a cutoff theory, it is
reasonable\ to assume\ translational invariance as a property of the vacuum
state and study the conditions for spontaneous symmetry breaking by an
effective potential analysis. An investigation of\ up to the two loop level
reveals that noncommutative effects can modify drastically the shape of the
effective potential. Under reasonable conditions, the nonplanar sector of the
theory can become dominant and induce symmetry breaking for values of the mass
and coupling constants not reached by the commutative counterpart.

\end{abstract}
\maketitle

\section{INTRODUCTION}

Quantum field theories with canonical noncommutativity of particle coordinates
became a subject of intensive investigation over the past years\ \cite{1,2}.
Characterized by the relation
\begin{equation}
\lbrack\hat{x}^{\mu},\hat{x}^{\nu}]=i\theta^{\mu\nu}, \label{1}%
\end{equation}
\ where $\theta^{\mu\nu}$ is an antisymmetric constant tensor, these theories
were shown to emerge as a natural approximation of string theory in its
low-energy limit \cite{4}, in D-brane physics and matrix theory \cite{5}, and
in the context of semiclassical gravity \cite{7}. They have also proved to be
applicable to condensed matter physics for the description of the quantum Hall
effect \cite{7.2} and superconductivity \cite{7.3}.

Among many interesting phenomena discovered in the study of noncommutative
quantum field theories, we quote Lorentz violation \cite{7.5},
nonlocality\ \cite{2}, and IR-UV mixing \cite{8}. The latter was shown to have
deep\ consequences, such as the modification of the conventional Wilsonian
picture of renormalization group flows in the very-low-momentum domain
\cite{2,9} and a change in the phase structure of a number of models
\cite{10}. From the experimental point of view, several attempts to establish
an empirical basis for\ noncommutativity\ are also under consideration (see,
e.g., \cite{11,11.5}).

Investigations were carried out to understand the role of noncommutativity in
the process of spontaneous symmetry breaking (SSB), most of them focused on
the $\varphi^{4}$model \cite{9,10,10.3,10.7,10.8,10.85}. Currently, there are
few investigations in (2+1)-dimensional $\varphi^{6}$ theory \cite{13,12}.
Reference \cite{12} is the only one on SSB, where an$\ O(N)$ model is analyzed
with\ emphasis on the issue of IR-UV mixing and renormalizability aspects.

The aim of the present work is to verify how noncommutativity may affect the
behavior\ of the $\varphi^{6}$ model as long as the conditions for SSB are
considered. We shall\ work with a cutoff field theory. Our motivation for this
comes mainly from three reasons. Until now, it has been unknown if the
noncommutative $\varphi^{6}$ model is renormalizable. The renormalizability of
noncommutative field theories is still under discussion (see,
e.g.,\cite{10,10.3,10.85,14,14.3} and references therein). The second
motivation is that, recently, numerical simulations \cite{10.9} are
corroborating results provided by noncommutative field theories based on a
Wilsonian approach\ \cite{10}, where a cutoff momentum is introduced.
Therefore, the cutoff models can provide interesting information on the vacuum
structure of field theories. A third reason comes from the fact that, after
all, quantum field theory is presently in the era of effective field theories
\cite{15}. The most successful theories, which are renormalizable, are
nowadays understood as low-energy approximations of a more fundamental theory
(perhaps not yet a field theory) and thus have their validity expected to
be\ limited up to an energy scale. Additional motivation, from the
phenomenological point of view, to consider the noncommutative models as
cutoff theories may be found in \cite{11.5}.

We shall restrict our considerations to the case of small $\theta$ and assume
translational symmetry as a property of the vacuum state. This assumption was
recently reassessed in the framework of noncommutative field theory (see,
e.g., \cite{10,10.7}). The argument behind these papers is that, as a
consequence of noncommutativity, the vacuum state of the models is no longer
realized for a constant $\varphi_{cl}\left(  x\right)  .$ When this is the
case,\ the effective potential is not a useful tool to analyze SSB. However,
as pointed out in \cite{10}, the particular case where $\theta$ is small, SSB
is expected to occur from a uniform-ordered to a disordered domain. Actually,
this prediction seems to be connected to the adoption of an explicit cutoff
for the theory, as we shall discuss later. It is interesting to quote that the
existence of a uniform-ordered domain was verified in a nonperturbative
calculation \cite{10.9} involving the three-dimensional $\varphi^{4}$ model.
Once a system is found in such a state, the minimum of the effective action is
realized for a constant $\varphi_{cl}$ and translational symmetry is a
property of the vacuum state.

The noncommutative effects relevant in the regime under consideration come
from the crossing of internal lines in the Feynman diagrams. Since the
finiteness of some of these diagrams is attributed to an effective cutoff
\cite{8}\ $\Lambda_{eff}\sim1/\sqrt{\theta}$, introduced by noncommutativity,
for $\theta$\ sufficiently small these diagrams may become dominant and
drastically modify the behavior of the field theories. Indeed, this is exactly
what will be shown to occur in this work.

This paper is organized as follows. Section II presents the noncommutative
version of the $\varphi^{6}$ model and path integral method employed for the
perturbative calculation of the effective potential. In Sec. III, the
effective potential is evaluated up to two-loop order. Section IV is devoted
to an analysis of the properties of the effective potential and the conditions
for\ SSB. Finally, in Sec. V, we end up with\ a general discussion and outlook.

\section{NONCOMMUTATIVE $\varphi^{6}$ MODEL}

\subsection{Background}

One important aspect concerning Eq. (\ref{1}) is that the transformation
properties of the indices $\mu$ and $\nu$ is\ not the same on\ the\ left- and
right-hand\ sides. The two sides carry Lorentz indices and transform under
coordinate changes, which characterize the observer (frame)\ Lorentz
transformations. However, the right-hand side is invariant under particle
Lorentz transformations, which do not act on $\theta^{\mu\nu}$ \cite{7.5}.
This explicitly shows how particle Lorentz symmetry is broken by
noncommutativity. Since $\theta^{\mu\nu}$ is Lorentz-observer covariant, we
can find a rigid orthogonal frame\ transformation $\tilde{x}=Lx$ that takes
the $\theta^{\mu\nu}$-matrix to its off-diagonal block\ form:\
\begin{equation}
L\left[
\begin{array}
[c]{ccc}%
0 & \theta^{01} & \theta^{02}\\
-\theta^{01} & 0 & \theta^{12}\\
-\theta^{02} & -\theta^{12} & 0
\end{array}
\right]  L^{T}=\left[
\begin{array}
[c]{ccc}%
0 & 0 & 0\\
0 & 0 & \theta\\
0 & -\theta & 0
\end{array}
\right]  \text{ ,} \label{2}%
\end{equation}
where $\theta^{2}=\left(  \theta^{01}\right)  ^{2}+\left(  \theta^{02}\right)
^{2}+\left(  \theta^{12}\right)  ^{2}$. Thus, in an intrinsic
three-dimensional world, with the appropriate choice for the physical frame,
noncommutativity of the space and time coordinates can be reduced to a pure
spatial noncommutativity, characterized by a unique parameter $\theta$.

Working with a noncommutative quantum\ field theory is equivalent to working
with a commutative quantum field theory by replacing the usual product in the
action by the star product of functions. The latter is defined as
\begin{equation}
\left(  f\star g\right)  (x)=\exp\left(  \frac{i}{2}\theta^{\mu\nu}%
\frac{\partial}{\partial\xi^{\mu}}\frac{\partial}{\partial\eta^{\nu}}\right)
f(x+\xi)g(x+\eta)\mid_{\xi=\eta=0,} \label{3}%
\end{equation}
and satisfies
\begin{align}
\int d^{3}x\text{ }\phi_{1}(x)\star\phi_{2}(x)...\phi_{n}(x)  &  =\int
\frac{d^{3}k_{1}}{\left(  2\pi\right)  ^{3}}...\frac{d^{3}k_{n}}{\left(
2\pi\right)  ^{3}}\left(  2\pi\right)  ^{3}\delta^{3}\left(  k_{1}%
+...+k_{n}\right) \nonumber\\
&  \times\exp\left(  -\frac{i}{2}\sum_{i<j}k_{i\mu}\theta^{\mu\nu}k_{j\nu
}\right)  \phi_{1}(k_{1})\phi_{2}(k_{2})...\phi_{n}(k_{n})\text{.} \label{4}%
\end{align}

\subsection{Formulation}

For the evaluation of the effective potential, we shall use path integral
methods \cite{40,41}. The Euclidean action of the noncommutative $\varphi^{6}$
theory is written as
\begin{equation}
S=\int\frac{d^{3}x}{(2\pi)^{3}}\left[  \frac{1}{2}\partial^{\mu}\tilde
{\varphi}\text{ }\partial_{\mu}\tilde{\varphi}+\frac{1}{2}m^{2}\tilde{\varphi
}^{2}+\frac{g}{4!}\tilde{\varphi}\star\tilde{\varphi}\star\tilde{\varphi}%
\star\tilde{\varphi}+\frac{f}{6!}\tilde{\varphi}\star\tilde{\varphi}%
\star\tilde{\varphi}\star\tilde{\varphi}\star\tilde{\varphi}\star
\tilde{\varphi}\right]  \text{.} \label{5}%
\end{equation}
Spontaneous symmetry breaking\ is introduced by allowing the quantum field,
$\mathbf{\tilde{\varphi}}$ to acquire a nonvanishing vacuum expectation
value,
\begin{equation}
\varphi_{cl}(x)=\langle0\mid\mathbf{\tilde{\varphi}}(x)\mid0\rangle_{J},
\label{6}%
\end{equation}
while the effective action $\Gamma\lbrack\varphi_{cl}]$\ develops an absolute
minimum for $\varphi_{cl}(x)=\langle0\mid\mathbf{\tilde{\varphi}}%
(x)\mid0\rangle_{J=0}.$

The perturbative loopwise expansion for the computation of $\Gamma
\lbrack\varphi_{cl}]$ via quantum corrections to Eq.\ (\ref{5})\ must
be\ performed around a stable vacuum. In order to do this, we substitute
$\tilde{\varphi}=\varphi_{cl}+\varphi$ in Eq.\ (\ref{5}) and expand around
$\varphi_{cl}$. This yields
\begin{align}
\mathcal{L}  &  =\frac{1}{2}\partial_{\mu}\varphi\partial^{\mu}\varphi
+\frac{1}{2}\left(  m^{2}+\frac{1}{2}g\varphi_{cl}^{2}+\frac{1}{24}%
f\varphi_{cl}^{4}\right)  \varphi^{2}+\frac{1}{3!}\left(  g\varphi_{cl}%
+\frac{1}{6}f\varphi_{cl}^{3}\right)  \varphi\star\varphi\star\varphi
\nonumber\\
&  +\frac{1}{4!}\left(  g+\frac{1}{2}f\varphi_{cl}^{2}\right)  \varphi
\star\varphi\star\varphi\star\varphi+\frac{1}{5!}\left(  f\varphi_{cl}\right)
\varphi\star\varphi\star\varphi\star\varphi\star\varphi+\frac{1}{6!}%
f\varphi\star\varphi\star\varphi\star\varphi\star\varphi\star\varphi,
\label{9}%
\end{align}
where the linear term in the fields disappears for its coefficient is\ the
classical field equation. To simplify the manipulations, we shall use the notation%

\begin{equation}
M^{2}=m^{2}+\frac{1}{2}g\varphi_{cl}^{2}+\frac{1}{24}f\varphi_{cl}^{4},\text{
\ \ \ \ }A=g\varphi_{cl}+\frac{1}{6}f\varphi_{cl}^{3},\text{ \ \ \ \ }%
B=g+\frac{1}{2}f\varphi_{cl}^{2},\text{ \ \ \ \ }C=f\varphi_{cl}\text{\ .}
\label{10}%
\end{equation}
The Euclidean effective action is given, in the functional formalism, by the
expression \cite{40,41}
\begin{equation}
\Gamma\lbrack\varphi_{cl}]=\int d^{3}x\mathcal{L}\left(  \varphi_{cl}\right)
+\frac{\hbar}{2}\ln\det\left(  \frac{\partial^{2}\mathcal{L}}{\partial
\varphi\partial\varphi}\right)  -\left(  \text{connected 1 PI diagrams}%
\right)  . \label{11}%
\end{equation}
As far as one assumes translational symmetry as a property of the model, the
vacuum\ structure of the theory can be determined by studying $\Gamma\left[
\varphi_{cl}\right]  $ for a constant $\varphi_{cl}=\phi.$ This amounts to an
analysis of the effective potential, which is defined as \cite{40,41}
\begin{align}
V\left(  \phi\right)   &  =\left.  \frac{\Gamma\left[  \varphi_{cl}\right]
}{\Omega}\right|  _{\varphi_{cl}=\phi}\nonumber\\
&  =V_{0}\left(  \phi\right)  +\frac{\hbar}{2}\int\frac{d^{3}k}{\left(
2\pi\right)  ^{3}}\ln\left(  k^{2}+M^{2}\right)  -\left\langle \exp\left(
-\frac{1}{\hbar}\int\frac{d^{3}k}{\left(  2\pi\right)  ^{3}}\mathcal{L}%
_{I}\left(  \varphi,\phi\right)  \right)  \right\rangle , \label{12}%
\end{align}
where $\Omega$ is the spacetime volume and $\mathcal{L}_{I}\left(
\varphi,\phi\right)  $ is the interaction part of the Lagrangian. The first
term in Eq. (\ref{12}) is the classical potential. It is given by the
interaction term of the action without the kinetic contributions originated
from the star product, that is,
\begin{equation}
V^{\left(  0\right)  }(\phi)=\frac{1}{2}m^{2}\phi^{2}+\frac{g}{4!}\phi
^{4}+\frac{f}{6!}\phi^{6}. \label{12.5}%
\end{equation}
The second term, containing the logarithm, is the contribution of all graphs
with one closed loop. The third one is the sum of the higher-order loop
corrections and is computed by taking the expectation value of
\begin{equation}
T\exp\left(  -\frac{1}{\hbar}\int\frac{d^{3}k}{\left(  2\pi\right)  ^{3}%
}\mathcal{L}_{I}\left(  \varphi,\phi\right)  \right)  \label{13}%
\end{equation}
using conventional Feynman rules.

The relevant vertices for the two-loop calculation of the effective potential
are\ depicted below:\unitlength2mm

\begin{picture}(33,12)
\put(15,0){\line(1,1){5}}
\put(20,5){\line(1,-1){5}}
\put(20,5){\line(0,1){5}}
\put(15,0){\vector(1,1){3}}
\put(25,0){\vector(-1,1){3}}
\put(20,10){\vector(0,-1){3}}
\put(17,0){$p_3$}
\put(22,0){$p_2$}
\put(21,8){$p_1$}
\put(27,5){$ -  A V(p_1,p_2,p_3)$}
\end{picture}
\begin{picture}(33,12)
\put(15,0){\line(1,1){10}}
\put(15,10){\line(1,-1){10}}
\put(15,0){\vector(1,1){3}}
\put(15,10){\vector(1,-1){3}}
\put(25,0){\vector(-1,1){3}}
\put(25,10){\vector(-1,-1){3}}
\put(17,0){$p_3$}
\put(17,10){$p_4$}
\put(22,0){$p_2$}
\put(22,10){$p_1$}
\put(27,5){$ -  B V(p_1,p_2,p_3,p_4)$}
\end{picture}

\section{EVALUATION OF THE EFFECTIVE\ POTENTIAL AT TWO-LOOP ORDER}

In this section we evaluate the effective potential for the $\varphi^{6}$
model\ up to two loops. Our calculations are rendered simpler by employing
analytic regularization. The regularizing factor will always be assumed as
finite. In what follows we shall work in units where $\hbar=1$.

The one-loop quantum contribution is evaluated using the trick
\begin{align}
V^{(1)}\left(  \phi\right)   &  =\frac{1}{2}\int\frac{d^{3}k}{\left(
2\pi\right)  ^{3}}\ln\left(  k^{2}+M^{2}\right)  =\frac{1}{2}\int d(M^{2}%
)\int\frac{d^{3}k}{\left(  2\pi\right)  ^{3}}\frac{1}{\left(  k^{2}%
+M^{2}\right)  ^{(1+\epsilon)}}\nonumber\\
&  =\frac{1}{8\pi^{3/2}}\frac{M^{\left(  -2\epsilon+3\right)  }}{3-2\epsilon
}\frac{\Gamma\left(  \epsilon-\frac{1}{2}\right)  }{\Gamma\left(
\epsilon+1\right)  }=-\frac{M^{3}}{12\pi}+O(\epsilon). \label{14}%
\end{align}
The\ two-loop correction is the sum of the double-bubble and\ sunset diagrams
in their planar and nonplanar versions. The planar double-bubble is given by \unitlength2.0mm%

\begin{align}
\begin{picture}(16.5,0) \put(9,0){\circle{4}} \put(13.2,0){\circle{4}}\end{picture}
&  =D_{1P}=-\frac{2}{3}\frac{B}{8}\int\frac{d^{3}k}{\left(  2\pi\right)  ^{3}%
}\frac{d^{3}p}{\left(  2\pi\right)  ^{3}}\frac{1}{\left(  k^{2}+M^{2}\right)
\left(  p^{2}+M^{2}\right)  }\nonumber\\
&  =-\frac{2}{3}\frac{B}{8}\frac{\pi^{3}}{\left(  2\pi\right)  ^{6}}\left(
\int\frac{e^{-\alpha M^{2}}}{\alpha^{3/2-\epsilon}}d\alpha\right)  ^{2}%
=-\frac{BM^{2}}{192\pi^{2}}+O(\epsilon). \label{15}%
\end{align}
The evaluation of the planar sunset can be done following the same procedure.
It yields%

\begin{align}
\begin{picture}(14.5,0) \put(11,0){\circle{4}} \put(9,0){\line(1,0 ){4}}\end{picture}
&  =D_{2P}=\frac{1}{2}\frac{A^{2}}{12}\int\frac{d^{3}k}{\left(  2\pi\right)
^{3}}\frac{d^{3}p}{\left(  2\pi\right)  ^{3}}\frac{1}{\left(  k^{2}%
+M^{2}\right)  \left(  p^{2}+M^{2}\right)  \left[  \left(  p+k\right)
^{2}+M^{2}\right]  }\nonumber\\
&  =\frac{1}{2}\frac{A^{2}}{12}\frac{\pi^{3}}{\left(  2\pi\right)  ^{6}}%
\int_{0}^{\infty}\alpha^{2}d\alpha\int_{0}^{1}dx\int_{0}^{1-x}dy\text{ }%
\frac{e^{-\alpha M^{2}}}{\alpha\left[  x\left(  1-x\right)  +y\left(
1-y\right)  -xy\right]  ^{1/2}}\nonumber\\
&  =\frac{A^{2}}{768\pi^{2}}\Gamma\left(  \epsilon\right)  \left(  \frac
{M}{\bar{\mu}}\right)  ^{-2\epsilon}=\frac{A^{2}}{768\pi^{2}\epsilon}%
-\frac{A^{2}}{768\pi^{2}}\ln\left(  \frac{M^{2}}{\bar{\mu}^{2}}e^{\gamma
}\right)  +O\left(  \epsilon\right)  , \label{16}%
\end{align}
where $\bar{\mu}$ is an arbitrary constant with mass dimension and $\gamma$ is
the Euler constant. In order that the effective potential do\ not be
explicitly dependent on the polar terms, we can perform the redefinitions
\begin{equation}
m_{R}^{2}=m^{2}-\frac{g^{2}}{384\pi^{2}\epsilon}\text{ \ \ , \ \ }%
g_{R}=g-\frac{fg}{96\pi^{2}\epsilon}\text{ \ \ , \ \ \ }f_{R}=f-\frac{5f^{2}%
}{192\pi^{2}\epsilon}. \label{16.5}%
\end{equation}
From now on, we shall assume the parameters of the model as defined by Eqs.
(\ref{16.5}) and omit the $R^{\prime}s$ from the notation. The expression for
$D_{2P}$ can therefore be written as
\begin{equation}
D_{2P}=-\frac{A^{2}}{768\pi^{2}}\ln\left(  \frac{M^{2}}{\mu^{2}}\right)  ,
\label{17}%
\end{equation}
where $\mu^{2}=\bar{\mu}^{2}e^{-\gamma}$.

The remaining contribution is given by the nonplanar versions of the diagrams
(\ref{15}) and (\ref{16}), which are read as
\begin{equation}
D_{1NP}=-\frac{1}{3}\frac{B}{8}\int\frac{d^{3}k}{\left(  2\pi\right)  ^{3}%
}\frac{d^{3}p}{\left(  2\pi\right)  ^{3}}\frac{e^{ik_{\mu}\theta^{\mu\nu
}p_{\nu}}}{\left(  k^{2}+M^{2}\right)  \left(  p^{2}+M^{2}\right)  }=-\frac
{1}{3}\frac{B}{8}I_{1}\label{18}%
\end{equation}
and
\begin{equation}
D_{2NP}=\frac{1}{2}\frac{A^{2}}{12}\int\frac{d^{3}k}{\left(  2\pi\right)
^{3}}\frac{d^{3}p}{\left(  2\pi\right)  ^{3}}\frac{e^{ik_{\mu}\theta^{\mu\nu
}p_{\nu}}}{\left(  k^{2}+M^{2}\right)  \left(  p^{2}+M^{2}\right)  \left[
\left(  p+k\right)  ^{2}+M^{2}\right]  }=\frac{1}{2}\frac{A^{2}}{12}%
I_{2}.\label{19}%
\end{equation}
By using the Feynman and Schwinger parametrizations, we can write $I_{1}$ as\
\begin{align}
I_{1} &  =\int_{0}^{1}dw\int_{0}^{\infty}d\alpha\int\frac{d^{3}kd^{3}%
p}{\left(  2\pi\right)  ^{6}}\alpha e^{ik_{\mu}\theta^{\mu\nu}p_{\nu}%
}e^{-\alpha\left[  \left(  k^{2}-p^{2}\right)  w+p^{2}+M^{2}\right]
}\nonumber\\
&  =\int_{0}^{1}dw\int_{0}^{\infty}d\alpha\int\frac{d^{3}ld^{3}p}{\left(
2\pi\right)  ^{6}}\alpha e^{-\alpha wl^{2}}e^{-\left(  1/4\alpha w\right)
\tilde{p}_{\mu}\tilde{p}^{\mu}}e^{-\alpha\left(  1-w\right)  p^{2}}e^{-\alpha
M^{2}},\label{20}%
\end{align}
where $\tilde{p}^{\mu}=\theta^{\mu\nu}p_{\nu}$ and $l^{\mu}=k^{\mu}-\left(
i/2\alpha w\right)  \theta^{\mu\nu}p_{\nu}$. In the frame defined by
Eq.\ (\ref{2}), $I_{1}$ is given by
\begin{align}
I_{1} &  =\int_{0}^{1}dw\int_{0}^{\infty}d\alpha\int\frac{d^{3}ld^{3}%
p}{\left(  2\pi\right)  ^{6}}\alpha e^{-\alpha wl^{2}}e^{-\left(  \theta
^{2}/4\alpha w\right)  \left(  p_{1}^{2}+p_{2}^{2}\right)  }e^{-\alpha\left(
1-w\right)  \left(  p_{0}^{2}+p_{1}^{2}+p_{2}^{2}\right)  }e^{-\alpha M^{2}%
}\nonumber\\
&  =\frac{\pi^{3}}{\left(  2\pi\right)  ^{6}}\int_{0}^{\infty}d\alpha\int
_{0}^{1}dw\frac{e^{-\alpha M^{2}}}{\left(  \alpha^{2}w^{3/2}\left(
1-w\right)  ^{3/2}+w^{1/2}\left(  1-w\right)  ^{1/2}\frac{\theta^{2}}%
{4}\right)  }\nonumber\\
&  =\frac{M^{2}}{32\pi}\frac{\left[  H_{0}\left(  \theta M^{2}\right)
-Y_{0}\left(  \theta M^{2}\right)  \right]  }{\theta M^{2}}\label{21}%
\end{align}
and, thus,
\begin{equation}
D_{1NP}=-\frac{BM^{2}}{768\pi}\frac{\left[  H_{0}\left(  \theta M^{2}\right)
-Y_{0}\left(  \theta M^{2}\right)  \right]  }{\theta M^{2}},\label{22}%
\end{equation}
where $H_{0}\left(  x\right)  $ is the Struve function and $Y_{0}\left(
x\right)  $ is a Bessel function of the\ second kind. Using Feynman and
Schwinger parametrizations, we can write $I_{2}$ as
\begin{align}
I_{2} &  =\int_{0}^{\infty}\alpha^{2}d\alpha\int_{0}^{1}dx\int_{0}^{1-x}%
dy\int\frac{d^{3}kd^{3}p}{\left(  2\pi\right)  ^{6}}\text{ }e^{-\alpha\left(
1-x\right)  p^{2}}e^{-\alpha\left(  1-y\right)  k^{2}}e^{-2\alpha\left(
1-x-y\right)  k_{\mu}p^{\mu}+ik_{\mu}\theta^{\mu\nu}p_{\nu}}e^{-\alpha M^{2}%
}\nonumber\\
&  =\int_{0}^{\infty}\alpha^{2}d\alpha\int_{0}^{1}dx\int_{0}^{1-x}dy\int
\frac{d^{3}ld^{3}p}{\left(  2\pi\right)  ^{6}}\text{ }e^{-\alpha\left(
1-x\right)  p^{2}}e^{-\alpha\left(  1-y\right)  l^{2}}e^{-\left[
1/4\alpha\left(  1-y\right)  \right]  \left[  -4\alpha^{2}\left(
1-x-y\right)  ^{2}p^{2}+\tilde{p}_{\mu}\tilde{p}^{\mu}\right]  }e^{-\alpha
M^{2}},\label{23}%
\end{align}
where $l^{\mu}=k^{\mu}-\left[  i/2\alpha\left(  1-y\right)  \right]  \left[
2i\left(  1-x-y\right)  \alpha p^{\mu}+\theta^{\mu\nu}p_{\nu}\right]  $. After
a simplification, Eq. (\ref{23}) is read as
\begin{align}
I_{2} &  =\frac{\pi^{3}}{\left(  2\pi\right)  ^{6}}\int_{0}^{\infty}d\beta
\int_{0}^{1}dx\int_{0}^{1-x}dy\text{ }\frac{\beta e^{-\beta}}{\left[  x\left(
1-x\right)  +y\left(  1-y\right)  -xy\right]  ^{1/2}}\nonumber\\
&  \hspace{7cm}\times\frac{1}{\left[  \left[  x\left(  1-x\right)  +y\left(
1-y\right)  -xy\right]  \beta^{2}+\frac{\theta^{2}M^{4}}{4}\right]
},\label{24}%
\end{align}
where $\beta=\alpha M^{2}$. This is difficult to solve analytically.
However,\ the contribution coming from this integral is finite and can be
neglected if compared to the contribution coming from $I_{1}$. The finiteness
of $I_{2}$ can be verified by considering the inequality
\begin{align}
I_{2}\left(  \theta M^{2}\right)   &  \leq\frac{\pi^{3}}{\left(  2\pi\right)
^{6}}\int_{0}^{\infty}d\alpha\int_{0}^{1}dx\frac{\alpha e^{-\alpha M^{2}}%
}{\left[  x\left(  1-x\right)  \right]  ^{1/2}}\frac{1}{\left[  x\left(
1-x\right)  \alpha^{2}+\frac{\theta^{2}}{4}\right]  }\nonumber\\
&  =-\frac{d}{dM^{2}}I_{1}=\frac{1}{32\pi}\left[  H_{1}\left(  \theta
M^{2}\right)  -Y_{1}\left(  \theta M^{2}\right)  \right]  -\frac{1}{16\pi^{2}%
}.\label{26}%
\end{align}
From Eq.\ (\ref{24}) it is possible to determine the dependence of $I_{2}$ on
$\theta$\ as
\[
I_{2}\sim\ln\left(  \theta^{2}M^{4}\right)  +O\left(  1\right)  ,
\]
and from Eq. (\ref{21}), we can write
\begin{equation}
I_{1}=-\frac{M^{2}}{32\pi^{2}\theta M^{2}}\ln\left(  \frac{\theta^{2}M^{4}}%
{4}e^{2\gamma}\right)  +O\left(  1\right)  .\label{27}%
\end{equation}
Thus, when $\theta$ is sufficiently small, $I_{1}$ is\ clearly the dominant
contribution, and $I_{2},$ as well as the one-loop correction, may be
discarded. Among all the contributions other than the one of\ $D_{1NP}$, we
shall keep only the classical piece, since it is the dominant\ one\ for large
$\phi.$ Therefore, in the regime of small $\theta$, the\ effective potential
is described by
\begin{equation}
V\left(  \phi\right)  =\frac{m^{2}}{2}\phi^{2}+\frac{g}{4!}\phi^{4}+\frac
{f}{6!}\phi^{6}+\frac{BM^{2}}{768\pi}\frac{H_{0}\left(  \theta M^{2}\right)
-Y_{0}\left(  \theta M^{2}\right)  }{\theta M^{2}}.\label{30}%
\end{equation}
Notice that, when $\theta\rightarrow0$, the nonplanar contribution behaves as
$\ln(\Lambda_{eff}^{2}/M^{2})\Lambda_{eff}^{2}/M^{2},$ rather than as
$\Lambda_{eff}^{2}/M^{2}.$ This may be attributed to the fact that
$\theta^{0i}=0,$ which means that\ the effective cutoff is absent in the
$p^{0}$ mode.

\section{ANALYSIS OF THE POTENTIAL}

\subsection{Tree-level Approximation}

Before considering the effective potential at two-loop order, it is
interesting to initially make some remarks about the tree approximation
\cite{43}. In order that the potential have an absolute minimum, it is
necessary that $f>0$.\ To perform the analysis of the vacuum, it is
interesting to consider the parameter $f$ fixed, and vary $m^{2}$ and $g$. The
possibilities for the variation of these parameters are (i) $m^{2}\geqslant0 $
and $g\geqslant0,$ (ii)$\ m^{2}\geqslant0$ and $g<0,$ (iii) $m^{2}<0$ and
$g\geqslant0,$ and\ (iv) $m^{2}<0$ and $g<0$.

Figure 1 shows the shape of the effective potential at the tree level for all
four possibilities in the thick solid lines of the graphs. In the first case
[Fig.1(a)], there is no SSB and the potential presents a minimum at the
origin. In the second, there is no SSB if $m^{2}>5g^{2}/8f$ [Fig.1(b)], but it
may occur if $m^{2}<5g^{2}/8f$ [Fig.1(c)], being characterized by the presence
of a local minimum at the origin and two global minima symmetrically disposed
around it. Cases (iii) and (iv) (Fig.1d)\ present SSB with a maximum at the
origin and two minima symmetrically disposed around it%

\begin{figure}
[ptb]
\begin{center}
\includegraphics[
height=5.93in,
width=6.2682in
]%
{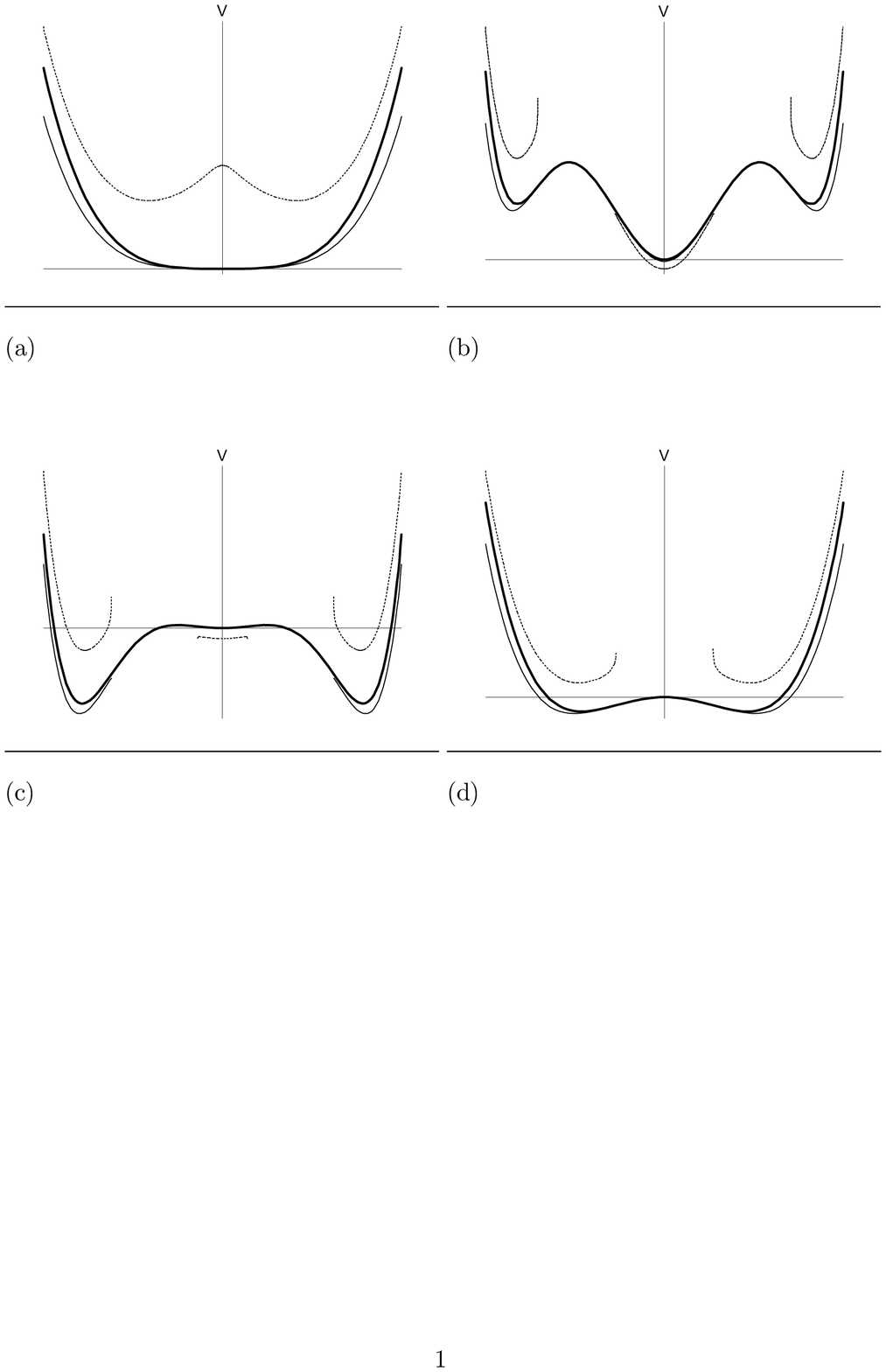}%
\caption{A typical picture of the effective potential $V\left(  \phi\right)
$\ at the tree level (thick solid lines) in contrast with its two-loop
commutative (thin solid lines) and noncommutative\ (dotted lines) versions.
(a) Case (i): $m^{2}\geqslant0$, $g\geqslant0$. (b) Case (ii): $m^{2}%
\geqslant0$, $g<0$ $(m^{2}>5g^{2}/8f)$. (c) Case (iii): $m^{2}\geqslant0$,
$g<0$ $(m^{2}<5g^{2}/8f)$. (d) Cases (iii), (iv): $m^{2}<0$, $g\geqslant0,$
$g<0.$}%
\end{center}
\end{figure}

\subsection{Two-loop potential}

A good evaluation of the modifications in the conditions for the SSB
introduced by the noncommutative effects may be carried out by comparing the
tree-level an two-loop corrected potentials in their noncommutative and
commutative versions. The latter is obtained by summing up Eqs.\ (\ref{12.5}),
(\ref{14}), and the planar diagrams (\ref{15}) and (\ref{16}) with their
weights$\ 2/3$ and $1/2$, respectively, redefined to be $1$. The resulting
commutative two-loop potential is
\begin{equation}
V\left(  \phi\right)  =\frac{m^{2}}{2}\phi^{2}+\frac{g}{4!}\phi^{4}+\frac
{f}{6!}\phi^{6}-\frac{M^{3}}{12\pi}+\frac{BM^{2}}{128\pi^{2}}+\frac{A^{2}%
}{384\pi^{2}}\ln\left(  \frac{M^{2}}{\mu^{2}}\right)  . \label{31}%
\end{equation}
In Fig. 1, it is possible to visualize the classical potential (thick solid
lines), the two-loop corrected noncommutative\ potential (doted lines),\ and
its two-loop corrected commutative counterpart (thin solid lines),\ for cases (i)-(iv).

In all cases, we see\ that the commutative two-loop corrections are manifest
through small shifts in the minima, moving them downwards with respect to
their positions at the tree level. The noncommutative two-loop corrections, on
the other hand, tend to modify drastically the original shape of the
potential. Notice that the dotted and thin solid lines are not continuous in
cases (ii)-(iv). This may be attributed to the fact that the quantum
corrections to the effective potential are not\ defined for the values of
$\phi$ for which $M^{2}<0$. Thus, at the\ two-loop level, it is not possible
to trace a complete picture of the potential with its minima in all cases.
Since the corrections to the tree-level approximation may become big enough to
displace the minima far from their original positions or alter their condition
of minima, it is not possible\ to determine with certainty the position of the
global minimum in cases (ii)-(iv). In connection with this, the most
interesting case to be considered is depicted in Fig. (1a), where the quantum
corrections are defined for all values of $\phi,$ and\ a SSB is shown to be
induced by noncommutativity effects. We shall, in what follows, restrict our
attention to this case.

The conditions for SSB in case (i) can be determined by studying the concavity
of the effective potential at the origin. Its second derivative at this point
is given by
\begin{equation}
\left.  \frac{d^{2}V}{d\phi^{2}}\right|  _{\phi=0}=m^{2}+\frac{m^{2}}%
{768\pi\theta m^{2}}\left\{  f\left[  H_{0}\left(  m^{2}\theta\right)
-Y_{0}\left(  m^{2}\theta\right)  \right]  -g^{2}\theta\left[  H_{1}\left(
m^{2}\theta\right)  -Y_{1}\left(  m^{2}\theta\right)  -\frac{2}{\pi}\right]
\right\}  . \label{33}%
\end{equation}
Notice that $m^{2}$ and the first term between brackets are both positive. The
second term containing $g^{2}\theta$, on the other hand, is negative and is
the one which tends to modify the concavity of the potential. Given a fixed
value for $\theta,$ it is easy to see, thanks to the behavior of the functions
$Y_{0}\left(  m^{2}\theta\right)  \sim\ln\left(  m^{2}\theta\right)  $ and
$Y_{1}\left(  m^{2}\theta\right)  \sim-1/m^{2}\theta$ near $m^{2}\theta=0$,
that\ for\ sufficiently large$\ g$, and $m^{2}$ and $f$ sufficiently\ small, a
reversal of concavity can occur.

Though in this work the formalism considered describes zero temperature, we
can speculate and regard the indirect influence of finite temperature via
physical parameters. A variation of temperature here could be regarded, for
example, as equivalent to a change in the mass parameter$\ m$. By varying
$m^{2},$ keeping $\theta$ fixed, it is easy to verify that the SSB generated
is a second-order phase transition.\footnote{We are considering the Landau
definition that a transition is first order if the order parameter $\left(
m^{2}\right)  $ is discontinuous at the transition point and second order if
it is continuous. Note that, for $n>2,$ this differs from the Ehrenfest
definition of an \textit{n}th order transition as the one in which
$\partial^{n}V/\partial\left(  m^{2}\right)  ^{n}$ is the lowest discontinuous
derivative.} To trace a picture of the dependence of the transition point
$m_{T}^{2}$ on the\ noncommutativity scale, we can solve $\left.  d^{2}%
V/d\phi^{2}\right\vert _{\phi=0}=0$ numerically for a wide range of values of
$\theta$.\ Figure 2 depicts $y=\log_{10}m_{T}^{2}$ as a function
of\ $x=\log_{10}\theta$ for given values of $f$ and $g.$%

\begin{figure}
[ptb]
\begin{center}
\includegraphics[
trim=0.049781in 0.000000in 0.000000in 0.050269in,
height=2.9706in,
width=3.96in
]%
{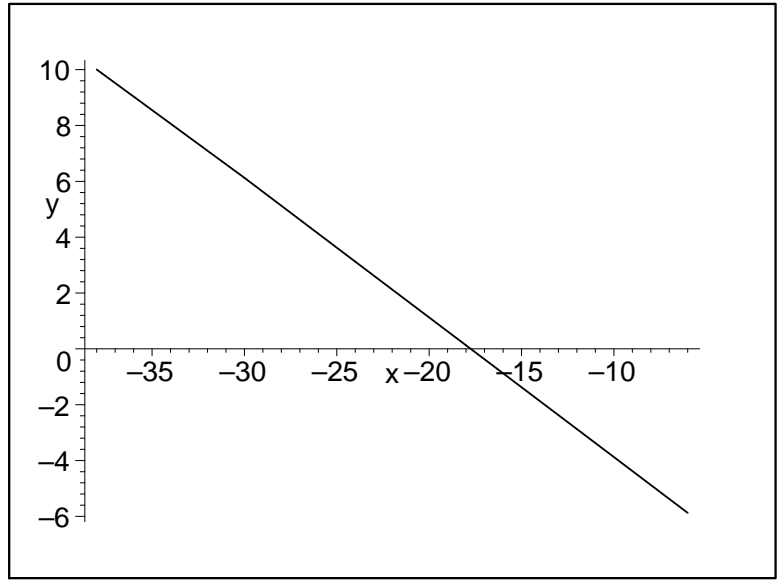}%
\caption{$\mathbf{y=\log}_{10}m_{T}^{2}$ $($GeV$^{2}),$ as a function of
$x=\log_{10}\theta$ $($GeV$^{-2})$ for $f=4.4\times10^{-27}$, and $\left.
g=8.2\times10^{-8}\text{ GeV}\right.  $.}%
\end{center}
\end{figure}

Since the values admissible for $\theta$ are small,\ an analytic approximation
for $m_{T}^{2}$\ can be obtained by expanding the right-hand side (RHS) of Eq.
(\ref{33}) in a series of $\theta$:
\begin{equation}
\left.  \frac{d^{2}V}{d\phi^{2}}\right|  _{\phi=0}=m^{2}-\frac{g^{2}}%
{384\pi^{2}m^{2}\theta}-\frac{f}{384\pi^{2}\theta}\ln\left(  \frac{m^{2}%
\theta}{2}e^{\gamma}\right)  +\frac{g^{2}}{384\pi^{2}}+\frac{f}{384\pi^{2}%
}m^{2}+O(\theta). \label{33.5}%
\end{equation}
Discarding the last three terms on the RHS of Eq. (\ref{33.5}), and solving
\ $\left.  d^{2}V/d\phi^{2}\right|  _{\phi=0}=0$, we obtain
\begin{equation}
m_{T}^{2}=\frac{g}{\sqrt{384\pi^{2}\theta}}, \label{34}%
\end{equation}
which reproduces the plot of Fig. 2 with great accuracy. In the ordinary
commutative $\varphi^{6}$ model, $m_{T}^{2}=0$. Thus, as long as
$g/\sqrt{\theta}$ is sufficiently big, the range of $m^{2}$ for which
noncommutativity-induced SSB occurs, $0<m^{2}<m_{T}^{2},$ can be very large.

\subsection{Considerations of translational invariance}

Up to now, we have assumed translational symmetry as a hypothesis. As pointed
out in \cite{10}, the SSB mechanism must involve the momentum modes where
$\Gamma^{(2)}$ is smallest. The presence of a global minimum of $\Gamma
^{(2)}(p_{0}^{2},p_{nc}^{2})$ away from $p_{\mu}=0$\ prevents condensation of
very-low-momentum modes and therefore breaks translational invariance.\ For
$\theta$\ sufficiently small, however, the minimum of $\Gamma^{(2)}$ here\ is
always at $p_{\mu}=0$, in a similar way as occurs in \cite{10}. The critical
value $\theta_{c}$ at which the minimum starts to move away from $p_{\mu}=0$
is estimated in a self-consistent way, for case (i), as follows.

Substituting $\phi=0$ in the vertex factors and accounting for the possible
crossings of the legs of the Feynman diagrams, it is possible to determine the
shape of the leading contributions when $\Gamma^{(2)}$ is evaluated\ up to
two-loop order as\footnote{We have discarded the two-loop corrections. For the
values fixed for $f$ and $g,$ it can be shown that their inclusion in the
calculation does not significantly\ modify the estimate for $\theta_{c}$.}
\begin{equation}
\Gamma^{(2)}(p_{0}^{2},p_{nc}^{2})\sim p_{0}^{2}+p_{nc}^{2}+g\Lambda_{eff},
\label{35}%
\end{equation}
where $p_{nc}^{2}=p_{1}^{2}+p_{2}^{2}$, $\Lambda_{eff}=1/\sqrt{\theta
^{2}p_{nc}^{2}+1/\Lambda^{2}},$ and $\Lambda$ is a UV cutoff. We have
neglected numerical coefficients and $p_{\mu}$-independent terms. In order
that the extremum equation $\nabla\Gamma^{(2)}(p_{0}^{2},p_{nc}^{2})=0$ admit
$p_{0}^{2}=p_{nc}^{2}=0$ as its unique solution, the following condition must
be satisfied:
\begin{equation}
\theta\Lambda^{2}<\left(  \theta\Lambda^{2}\right)  _{c}\sim1/(g/\Lambda
)^{1/2}. \label{36}%
\end{equation}
It is easy to check that the corresponding extremum is a minimum. We can fix
$\Lambda$ and write Eq.\ (\ref{36}) as $\theta<\theta_{c}\sim1/(g\Lambda
^{3})^{1/2}$ or fix $\theta$ and write Eq.\ (\ref{36}) as $\Lambda<\Lambda
_{c}\sim1/(g\theta^{2})^{1/3}.$ Let us fix some values for $\theta$ and
compute the associated $\Lambda_{c}$ first. For a noncommutativity (NC)
parameter around $\theta\sim10^{-24}$ GeV$^{-2}$ (an experimental bound found
by Anisimov \textit{et al}. \cite{11.5} from NCQED\ with $\Lambda=1$ TeV), we
have $\Lambda_{c}\sim10^{18}$ GeV, near the Planck scale. On the other hand,
if $\theta\sim10^{-30}$ GeV$^{-2}$, which is the most stringent experimental
bound proposed for this parameter (see Mocioiu \textit{et al}. \cite{11}), we
have $\Lambda_{c}\sim10^{22}$ GeV, which is beyond the Planck scale. For
$\theta$ even smaller, the corresponding $\Lambda_{c}$ is yet larger. Since a
reasonable value for $\Lambda$ is several times smallest than the Planck
scale, the condition $\theta<\theta_{c}\sim1/(g\Lambda^{3})^{1/2}$ can be
easily satisfied. For $\Lambda=1$ TeV, e.g., we would have $\theta_{c}%
\sim10^{-1}$ GeV$^{-2}\gg10^{-24}$ GeV$^{-2}$.

\section{DISCUSSION AND OUTLOOK}

In this work, we have carried out the study of aspects of SSB for the
noncommutative $\varphi^{6}$ model. Our main goal was the investigation of the
relevance of noncommutativity effects when $\theta$ is small. An analysis
carried out at two-loop order in perturbation theory\ revealed that
the\ noncommutative corrections to the effective potential\ are dominated by
the nonplanar contributions. These are very different from the corresponding
planar counterparts, which, except for weight factors, are functionally
similar to the ordinary commutative ones. The bad behavior of the nonplanar
diagrams in the $\theta\rightarrow0$ limit\ can be interpreted as a
consequence of the removal of the effective cutoff $\Lambda_{eff}\sim
1/\sqrt{\theta},$ naturally introduced by noncommutative geometry. This is
exactly what makes the study of the nonplanar contributions in the
small-$\theta$\ limit especially interesting,\ since they can drastically
modify the shape of the effective potential. Indeed, this was the general
conclusion of our qualitative analysis in all cases covered.

As the main result of our investigation, we found that, at the two-loop level,
noncommutativity effects can induce a SSB for positive values of the mass
parameter and the coupling constants. The conditions for this to occur were
determined in a quantitative analysis, which provided an analytical expression
for the square of the mass parameter, $m_{T}^{2}$, where the SSB takes place.
For\ the values fixed for the coupling constants $g$ and $f,$ the mass channel
for which the noncommutativity-induced SSB occurs, $0<m^{2}<m_{T}^{2}$, was
shown to comprise a reasonable range of values and to become large as long as
$\theta$ gets smaller. Other choices for $g$, $f$ may yet expand the
possibilities for the mass channel.

For the fixed\ values for $g$ and $f$, we showed that, for acceptable $\theta
$\ and reasonable values of $\Lambda,$ translational symmetry can be assumed
as a property of the vacuum state. If one considers $\Lambda$ as a parameter
to vary, the corresponding $\theta_{c}(\Lambda)$ is a decreasing function. In
the continuum limit (in the case it makes sense), $\theta_{c}=0 $ and the
translational invariant phase ceases to exist. The same occurs if one
considers, e.g., the expression for $\theta_{c}$ presented for the
$\varphi^{4}$ model in \cite{10}: $\theta_{c}\sim1/(g\Lambda^{2}) $, where
$g^{2}$ is the coupling constant. The existence of a translational invariant
phase\ seems to be a\ common\ feature of the cutoff models, where, for
$\theta$ sufficiently small, $\Gamma^{(2)}(p_{\mu})$ admits its global minimum
at $p_{\mu}=0$. In a previous work \cite{14} it was argued that, for the
noncommutative scalar field theories to be renormalizable, it would be
necessary to relax the hypothesis of translational invariance. Perhaps
translational invariance for small $\theta$ is therefore a property only of
the cutoff models.

Beyond the\ natural interest that is motivating the investigation of
noncommutativity from the theoretical point of view, knowledge of how it can
modify the shape of the effective potential for scalar theories could be of
great relevance, for example, in cosmology. The\ analysis of effective
potentials\ plays a fundamental role in the study of inflationary universe
models \cite{45}. Since the $\theta^{ij}$'s, if nonvanishing, should be very
small, the results obtained in this work may be useful as clues for a possible
application of noncommutativity as a mechanism for generating a symmetry
breaking in\ the inflationary scenario. An example of how noncommutative field
theory may be applied\ to inflationary cosmology may be found in \cite{46}.

\section*{Acknowledgments}

The author is greatly indebted to Jos\'{e} Hela\"{y}el-Neto for discussions
and all corrections on this manuscript. He also acknowledges Sebasti\~{a}o
Alves Dias,\ Wolfgang Bietenholz, and Victor Rivelles for suggestions and a
critical reading of earlier versions of the manuscript.\ This work was
financially supported by CAPES.


\begin{thebibliography}{99}                                                                                               %


\bibitem {1}R. J. Szabo, Phys. Rep. \textbf{378,} 207 (2003).

\bibitem {2}M. R. Douglas and N. A.~Nekrasov, Rev. Mod. Phys. \textbf{73,} 977 (2002).

\bibitem {4}N. Seiberg and E. Witten, J. High Energy Phys. \textbf{09,} 032 (1999).

\bibitem {5}A.~Connes, M. R. Douglas and A.~Schwarz, J. High Energy Phys.
\textbf{02},003 (1998) 003;

M \ R. Douglas and C. M. Hull, J. High Energy Phys. \textbf{02}, 008 (1998).

\bibitem {7}S. Doplicher, K. Fredenhagen and J. E. Roberts, Phys. Lett.
\textbf{B 331,} 39 (1994); Comm. Math. Phys. \textbf{172,} 187 (1995).

\bibitem {7.2}A. P. Polychronakos, J. High Energy Phys. \textbf{04}, 011 (2001);

L. Susskind, \textquotedblleft The Quantum Hall Fluid and Noncommutative
Chern-Simons Theory\textquotedblright, hep-th/0101029.

\bibitem {7.3}J. Yeo, C. Rim and K. Moon, J. Phys. \textbf{A 37}, L39 (2004).

\bibitem {7.5}S.~Carroll, J.~Harvey, V. A.~Kostelecky, C. D.~Lane and
T.~Okamoto, Phys. Rev. Lett. \textbf{87}, 141601 (2001).

\bibitem {8}S.~Minwalla, M.~Van~Raamsdonk and N.~Seiberg, J. High Energy Phys.
\textbf{02}, 020 (2000).

\bibitem {9}B. A. Campbell and A. Kaminsky,\ Nucl. Phys. \textbf{B 581,} 240 (2000).

\bibitem {10}S. S. Gubser and S. L.Sondhi, Nucl. Phys. \textbf{B 605}, 395 (2001).

\bibitem {11}I. Mocioiu, M. Pospelov and R. Roiban, Phys. Lett. \textbf{B
489}, 390 (2000);

G.~Amelino-Camelia, L.~Doplicher, S.-K. Nam and Y.-S. Seo, Phys. Rev.
\textbf{D 67} 085008 (2003);

I.~Hinchliffe and N.~Kersting, Int. J. Mod. Phys. \textbf{A19}, 179 (2004);

G.~Amelino-Camelia, G.~Mandanici and K.~Yoshida, J. High Energy Phys.
\textbf{01}, 037 (2004).

\bibitem {11.5}A. Anisimov, T. Banks, M. Dine and M. Graesser, Phys. Rev.
\textbf{D 65}, 085032 (2002);

I.~Hinchliffe and N.~Kersting, Phys. Rev.\ \textbf{D 64}, 116007 (2001).

\bibitem {10.3}Guang-Hong Chen and Yong-Shi Wu, Nucl. Phys. \textbf{B 622,}
189 (2002).

\bibitem {10.7}G. Mandanici, Int. J. Mod. Phys. \textbf{A19,} 3541 (2004).

\bibitem {10.8}W. H. Huang, Phys. Lett. \textbf{B 496,} 206 (2000).

\bibitem {10.85}F. Ruiz Ruiz, Nucl. Phys. \textbf{B 637},143 (2002).

\bibitem {13}Y. Kiem, C. Kim and Y. Kim, Phys. Lett. \textbf{B 507,} 207 (2001).

\bibitem {12}H. O. Girotti, M. Gomes, A. Yu. Petrov, V. O. Rivelles and A. J.
da Silva, Phys. Rev. \textbf{D 67}, 125003 (2003).

\bibitem {14}S. Sarkar, J. High Energy Phys.\textbf{ 02}, 030 (2002).

\bibitem {14.3}I.~Chepelev and R.~Roiban, J. High Energy Phys. \textbf{03},
001 (2001).

\bibitem {10.9}W. Bietenholz, F. Hofheinz and J. Nishimura, Fortsch. Phys.
\textbf{51}, 745 (2003).

\bibitem {15}S. Weinberg, The Quantum Theory of Fields, Cambridge University
Press, 1996.

\bibitem {40}R. Jackiw, Phys. Rev. \textbf{D 9}, 1686 (1974).

\bibitem {41}M. Peskin and D. Schroeder, An Introduction to Quantum Field
Theory, Perseus Books, Cambridge, Massashussets, 1995.

\bibitem {43}K. Huang, Statistical Mechanics, John Wiley \& Sons, Second
Edition, 1987.

\bibitem {45}A. D. Linde, Rep. Prog. Phys. \textbf{42 }(1979).

E. W. Kolb and M. S. Turner, The Early Universe, Perseus Publishing, 1994.

\bibitem {46}C.-S. Chu, B. R. Greene and G. Shiu, Mod. Phys. Lett. \textbf{A
16},\ 2231 (2001).
\end{thebibliography}
\end{document}